\documentclass[12pt]{article}
\usepackage{amsmath, amssymb}
\def\a{\alpha}

\def\t{\tau}

\def\8{\infty}
\def\sgn{\mbox{{\rm sgn}}}
\def\la{\langle}
\def\ra{\rangle}
\newtheorem {thm}{Theorem}[section]
\newtheorem {lem}[thm]{Lemma}
\newtheorem {cor}[thm]{Corollary}

\begin{document}
\title{Difference system for Selberg correlation integrals}
\author{Peter J. Forrester$^*$ and Masahiko Ito$^\dagger$}
\date{}
\maketitle

\noindent
\thanks{\small $^*$ Department of Mathematics and Statistics, University of Melbourne,
Victoria 3010, Australia;\\ $^\dagger$ School of Science and Technology for future life,
Tokyo Denki University,
Tokyo 101-8457, Japan}
\begin{abstract}
\noindent
%\textcolor{red}{
The Selberg correlation integrals are averages of the products $\prod_{s=1}^m\prod_{l=1}^n (x_s - z_l)^{\mu_s}$ 
with respect to the Selberg density. Our interest is in the case $m=1$, $\mu_1 = \mu$, when
 this corresponds to the $\mu$-th moment of the
corresponding characteristic polynomial. We give the explicit form of a $(n+1) \times (n+1)$ matrix linear
difference system in the variable $\mu$ which determines the average, and we give the Gauss decomposition
of the corresponding $(n+1) \times (n+1)$ matrix. For $\mu$ a positive integer the difference system can be used to
efficiently compute the power series defined by this average.
%}
\end{abstract}
\section{Introduction}
The probability density function (PDF) on $z_l \in [0,1]$ $(l=1,\dots,N)$
\begin{equation}\label{1.1}
{1 \over S_n(\alpha_1,\alpha_2,\tau)} \prod_{i=1}^n z_i^{\alpha_1 - 1} (1 - z_i)^{\alpha_2 - 1}
\prod_{1 \le j < k \le n} |z_j - z_k|^{2 \tau},
\end{equation}
where
\begin{eqnarray}\label{1.1a}
S_n(\alpha_1,\alpha_2,\tau) & := & 
\int_{[0,1]^n} \prod_{i=1}^n z_i^{\alpha_1 - 1} (1 - z_i)^{\alpha_2 - 1}
\prod_{1 \le j < k \le n} |z_j - z_k|^{2 \tau} \, dz_1 \cdots dz_n \nonumber \\
& = & 
 \prod_{j=0}^{n-1} {\Gamma (\alpha_1  + j\tau)
\Gamma (\alpha_2 + j\tau)\Gamma(1+(j+1)\tau) \over
\Gamma (\alpha_1 + \alpha_2  + (n + j-1)\tau) \Gamma (1 + \tau )},
\label{3.2}
\end{eqnarray}
is the Selberg integral, plays a fundamental role in the theories of random matrices and
Calogero-Sutherland quantum many body systems (see e.g.~\cite{Fo02}). In random matrix theory,
(\ref{1.1}) with $(\alpha_1,\alpha_2,\tau) = (a+1,b+1,\beta/2)$ defines the Jacobi
$\beta$-ensemble. For $\beta = 1,2$ and 4, and certain $a,b$, this can be realized as the
eigenvalue PDF occurring in the analysis of correlation coefficients associated with
Gaussian data sets, and also as the singular values of sub-matrices formed from various
unitary matrices chosen according to the Haar measure. For general $\alpha_1,\alpha_2, \tau > 0$
there are constructions of (\ref{1.1}) relating to similarity reductions of unitary matrices to
Hessenberg form \cite{KN04}, to block diagonal form \cite{ES06a}, and to the generalized eigenvalue problem
for certain tridiagonal matrices \cite{FR02b,Zh99}.

Upon the change of variables $z_i = \sin^2 \phi_i$, $0 < \phi_i < \pi/2$ $(i=1,\dots,n)$,
the PDF~(\ref{1.1}) becomes proportional to
$$
\prod_{i=1}^n (\sin^2 \phi_i)^{\alpha_1'} (\cos^2 \phi_i)^{\alpha_2'}
\prod_{1 \le j < k \le n} | \sin^2 \phi_j - \sin^2 \phi_k|^{2 \tau}
$$
with $\alpha_1' = \alpha_1 - 1/2$, $\alpha_2' = \alpha_2 - 1/2$.
As such it is the absolute value squared of the ground state wave function for the
$BC$-type Calogero-Sutherland Schr\"odinger operator \cite[Eq.(11.55)]{Fo02}
\begin{eqnarray*}
&& - \sum_{j=1}^n {\partial^2 \over \partial \phi^2} +
\sum_{j=1}^n \Big ( {\alpha_1'\tau(\alpha_1'\tau - 1) \over \sin^2 \phi_j} +
{\alpha_2'\tau(\alpha_2'\tau - 1) \over \cos^2 \phi_j} \Big ) \\
&& \qquad \qquad + 2 \tau (\tau - 1)
\sum_{1 \le j < k \le n}
\Big ( {1 \over \sin^2(\phi_j - \phi_k)} + {1 \over \sin^2(\phi_j + \phi_k)} \Big ).
\end{eqnarray*}

It is well known that (\ref{1.1}) exhibits many remarkable integrability properties. One is the
gamma function form of the normalization (\ref{1.1a}). Another is that the family of averages 
associated with (\ref{1.1})
\begin{equation}\label{2.1}
\int_{[0,x]^p} \int_{[0,1]^{n-p}} \prod_{i=1}^n z_i^{\alpha - 1}
(1 - z_i)^{\beta - 1} |x - z_i|^{\mu } \prod_{1 \le j < k \le n}
|z_j - z_k|^{2 \tau} \, dz_1 \cdots dz_n
\end{equation}
(we have set $\alpha_1 = \alpha$, $\alpha_2 = \beta$)
can be characterized in terms of a certain differential-difference equation \cite{Fo93},
equivalent to a $(n+1) \times (n+1)$ matrix Fuchsian differential equation \cite{FW07p,Mi07}.
With $p=0$ and $\mu = 2 \tau $, (\ref{2.1}) is simply related to the one-point
density implied by (\ref{1.1}), and the differential-difference equation was used in
\cite{Fo93} to compute the polynomial in $x$ specified by (\ref{2.1}) in the case
$\tau \in \mathbb Z_+$ (the polynomial is of degree $2 \tau n$ so for practical
purposes $\tau n$ cannot be too large).

In the case $\tau = 1$ (\ref{2.1}) can be calculated in terms of the solution of the
Painlev\'e VI non-linear differential equation in $\sigma$ form \cite{FW04}. It is also
revealed in \cite{FW04} that the $\sigma$-function associated with (\ref{2.1})
satisfies a third order non-linear difference equation for integer shifts
in the variable $\mu$, while (\ref{2.1}) itself can be computed by a recurrence scheme
based on the discrete Painlev\'e V equation. This can be understood from the
viewpoint of a more general theory relating to isomonodromic deformations of
linear difference systems \cite{Bo04}.

It is the objective of this work to provide a $(n+1) \times (n+1)$ matrix
linear difference system for integer shifts of
the variables $\alpha$, $\beta$ or $\mu$ (the latter restricted to cases that
$(x - z_i) |x - z_i|^{\mu } = \pm |x - z_i|^{\mu + 1}$ for some sign $\pm$, or
alternatively to twice integer shifts) in the integrals
(\ref{2.1}) with $\tau > 0$. This will be used to provide an alternative method to compute the
polynomial in $x$ specified by (\ref{2.1}) in the case $p=0$, $\mu - 1$ even.

A precise formulation of the family of Selberg correlation integrals to be
studied is given in Section 2, along with a statement of our
result for the explicit form of the difference system. In Section 3 we introduce a certain family of
interpolation polynomials, and we state three term relations satisfied by the corresponding
generalized Selberg integrals. The three term relations are shown to imply the difference system. We
give their  proof  in Section 4. In Section 5 it is shown how to use the difference
system to compute (\ref{2.1}) in the polynomial case. Furthermore, we specify applied studies in
random matrix theory to which our computations have relevance, and we make note too of the
wider problem of characterizing correlation functions in statistical mechanical problems in terms
of differential or difference equations.

\section{Definitions and main result}
We begin with some definitions.
We introduce the notation $\Phi(z)$, $z := (z_1,\dots,z_n)$, to denote the generalization of the
integrand (\ref{2.1}),
\begin{equation}\label{3a}
\Phi(z) := \prod_{i=1}^n | x_1 - z_i|^{\alpha_1 - 1} | x_2 - z_i|^{\alpha_2 - 1} 
| x_3 - z_i|^{\alpha_3 - 1} \prod_{1 \le j < k \le n} |z_j - z_k|^{2 \tau}.
\end{equation}
The parameters $\alpha_1, \alpha_2, \alpha_3$ are assumed restricted to domains for which it is
possible to specify a region $\Delta \subset \mathbb R^n$ with the property $\Phi(z)$ vanishes on
the boundary $\partial \Delta$ of $\Delta$. For example, if $0 < x < 1$ and
${\rm Re} (\alpha_1)$, ${\rm Re} (\alpha_2)$, ${\rm Re} (\alpha_3) > 0$, we can specify
\begin{equation}\label{4.1}
\Delta = \Delta_p = [0,x]^p \times [0,1]^{n-p} \qquad (p=0,1,\dots,n).
\end{equation}
For rational functions $\phi(z)$ bounded on $\Delta$ we define
\begin{equation}\label{4.0}
\langle \phi \rangle := \int_{\Delta} \phi(z) \Phi(z) \, dz_1 \cdots dz_n,
\end{equation}
and we use this notation in turn to specify $T_{\alpha_j} $ according to
\begin{equation}\label{T1}
T_{\alpha_j} \langle \phi \rangle = \Big \langle \prod_{i=1}^n (z_i - x_j) \phi \Big \rangle.
\end{equation}
Note that in the cases that $(z_i - x_j) | z_i - x_j|^{\alpha_j - 1} = \pm |z_i - x_j|^{\alpha_j}$
for some sign $\pm$, $T_{\alpha_j}$ corresponds to incrementing $\alpha_j$ by 1, and
independent of this requirement, $T_{\alpha_j}^2$ corresponds to incrementing $\alpha_j$ by 2.
Our goal is to identify polynomials $\{\varphi_i(z)\}_{i=0,1,\dots,n+1}$ such that
$\{T_{\alpha_1} \langle \varphi_i \rangle \}_{i=0,1,\dots,n+1}$ is linearly related to
$\{\langle \varphi_i \rangle \}_{i=0,1,\dots,n+1}$. 

For this purpose we take inspiration from the work of Aomoto \cite{Ao75,Ao87}. With
$\Phi^*(z)$ denoting $\Phi(z)$ specialized to $x_1 = 0$, $x_2 = 1$, $\alpha_3 = 1$
so that
\begin{equation}\label{5a}
\Phi^*(z) := \prod_{i=1}^n z_i^{\alpha_1 - 1} (1 - z_i)^{\alpha_2 - 1}
\prod_{1 \le j < k \le n} |z_j - z_k|^{2 \tau}
\end{equation}
and with
\begin{equation}\label{5b}
\langle \phi \rangle^* := \int_{[0,1]^n} \phi(z) \Phi^*(z) \, dz_1 \cdots dz_n
\end{equation}
it was proved in \cite{Ao87} that
$$
\Big \langle \prod_{l=1}^{i+1} z_l \Big \rangle^* =
{\alpha_1 + (n-i-1) \tau \over \alpha_1 + \alpha_2 + (2n - i - 2) \tau}
\Big \langle \prod_{l=1}^{i} z_l \Big \rangle^*.
$$
Since
$
S_n(\a_1+1,\a_2,\t)=\int z_1\cdots z_{n}\,\Phi^*(z) \, dz_1\cdots dz_n,
$
by iterating this
we immediately have the difference equation for the Selberg integral of (\ref{1.1a}),
\begin{equation}\label{Sa}
S(\a_1+1,\a_2,\t)=S(\a_1,\a_2,\t)\prod_{i=1}^n\frac{\a_1+(n-i)\t}{
\a_1+\a_2+(2n-i-1)\t},
\end{equation}
which in turn can be used to deduce the Gamma function evaluation given in (\ref{1.1a}).
Thus we learn that in the case of the Selberg integral $T_{\alpha_1} \langle 1 \rangle^* $ is
linearly related to $ \langle 1 \rangle^*$, and moreover we note that to derive the linear relation
use was made of the auxilary polynomials $\{\prod_{l=1}^i z_l \}$ --- referred to as interpolation
polynomials for the role they play in the calculation.

We will follow the same general strategy in relation to the integrals (\ref{4.0}).
Thus a family of interpolation polynomials, $\{\varphi_{i,j}(z)\}$, will be introduced so
as to determine the polynomials $\{\varphi_i(z)\}_{i=0,1,\dots,n+1}$ forming  
a polynomial basis of the difference system associated with the shift
$\alpha_1 \mapsto \alpha_1 + 1$ in the integrals (\ref{4.0}).
We know from \cite{Ao75,Ao87} that the main tool in determining these polynomials
is the vanishing of a certain class of averages (\ref{4.0}).

\begin{lem}
\label{lem:nabla}
For $k=1,2,\ldots, n$ let
\begin{eqnarray}
(\nabla_{\!k}\,\phi)(z)&:=&\frac{\partial\phi}{\partial z_k}(z)+\frac{\phi(z)}{\Phi(z)}\frac{\partial\Phi}{\partial z_k}(z) \nonumber \\
&=&\frac{\partial\phi}{\partial z_k}(z)
+\Big(
-\frac{\a_1-1}{x_1-z_k}-\frac{\a_2-1}{x_2-z_k}-\frac{\a_3-1}{x_3-z_k}
+\sum_{1\le l\le n\atop l\ne k}\frac{2\t}{z_k-z_l}\Big)\phi(z).
\label{eq:nabla}
\end{eqnarray}
We have $\la\nabla_{\!k}\,\phi\ra=0$.
\end{lem}
{\bf Proof.}
By definition,
$$
\la\nabla_{\!k}\,\phi\ra=\int_{\Delta} \Phi(z)\nabla_{\!k}\,\phi(z)dz_1\cdots dz_n
=\int_{\Delta}\frac{\partial}{\partial z_k}\Big(\phi(z)\Phi(z)\Big)dz_1\cdots dz_n=0
$$
if $\phi(z) \Phi(z)$ vanishes on the boundary $\partial \Delta$ of $\Delta$, which we have
assumed. \hfill $\Box$\\

However, 
for purposes of presentation, rather than to start with the interpolation polynomials,
it is convenient to immediately
present our findings for the explicit form of the polynomials
$\{\varphi_i(z)\}_{i=0,1,\dots,n+1}$ and the corresponding difference system.

\begin{thm}\label{thm2.2}
Write
\begin{equation}\label{5}
\varphi_i(z):=\underbrace{ (x_2-z_1)\cdots(x_2-z_{n-i})}_{n-i}
\times\underbrace{ (x_3-z_{n-i+1})\cdots(x_3-z_n)}_{i}
\qquad i=0,1,\ldots, n.
\end{equation}
We have
\begin{equation}\label{9}
T_{\a_1}(\la \varphi_0\ra,\la \varphi_1\ra,\ldots,\la \varphi_n\ra)
=(\la \varphi_0\ra,\la \varphi_1\ra,\ldots,\la \varphi_n\ra)A
\end{equation}
where  $A=LDU$ with
\[L=
\left(\!\!
\begin{array}{cccc}
l_{00} &   &  & \\
l_{10} & l_{11} & & \\
\cdots & \cdots & \cdots & \\
l_{n0} & l_{n1} & \cdots & l_{nn}
\end{array}
\!\!\right),
\quad
D=
\left(\!\!
\begin{array}{cccc}
 d_{0} & & & \\
  & d_{1}& & \\
  &   & \cdots &\\
 & & & d_{n}
\end{array}
\!\!\right),
\quad
U=
\left(\!\!
\begin{array}{cccc}
 u_{00} & u_{01}  & \cdots & u_{0n}\\
  & u_{11}& \cdots & u_{1n}\\
  &   & \cdots & \cdots\\
 & & & u_{nn}
\end{array}
\!\!\right)
.
\]
All entries in $L,D,U$ not explicitly shown are zero, while for the non-zero entries we have
\begin{eqnarray}\label{9a}
l_{ij}&=&(-1)^{i-j}{n-j\choose n-i}
\frac{(\a_2+j\t;\t)_{i-j}}{\big(\a_1+\a_2+2j\t;\t\big)_{i-j}}
\bigg(\frac{x_2-x_1}{x_3-x_1}\bigg)^{\!\! i-j},
\nonumber \\[7pt]
d_{j}&=&\frac{(\a_1;\t)_j\big(\a_1+\a_2+2j\t;\t\big)_{n-j}(x_2-x_1)^j(x_3-x_1)^{n-j}}
{(\a_1+\a_2+(j-1)\t;\t\big)_j\big(\a_1+\a_2+\a_3+(n+j-1)\t;\t\big)_{n-j}},
\nonumber \\[7pt]
u_{ij}&=&
(-1)^{j-i}{j\choose i}
\frac{
\big(\a_3+(n-j)\t;\t\big)_{j-i}}{\big(\a_1+\a_2+2i\t;\t\big)_{j-i}},
\end{eqnarray}
where $(x;\t)_0=1$ and
$(x;\t)_i:=x(x+\t)(x+2\t)\cdots(x+(i-1)\t)$ for $i=1,2,\ldots$.
\end{thm}

The proof of this result will be given in Section 3.

In Theorem \ref{thm2.2} the matrix $A$ is given in terms of its Gauss $LU$ decomposition.
The symmetry of (\ref{3a}) under the interchange $(x_2,\alpha_2) \mapsto (x_3,\alpha_3)$ allows
for $A$ also to be written in terms of its $UL$ Gauss decomposition. To see this, let
$\bar{L}$, $\bar{D}$ and $\bar{U}$ be the matrices $L,D$ and $U$ after this interchange.
Since
$$
(\la \varphi_n\ra,\la \varphi_{n-1}\ra,\ldots,\la \varphi_0\ra)=(\la \varphi_0\ra,\la \varphi_1\ra,\ldots,\la \varphi_n\ra)J,
$$
where
$$
J=
\left(
\begin{array}{cccc}
 &   &  &1 \\
 &  &1 & \\
 & \cdots & & \\
1 & &  &
\end{array}
\right),
$$
(\ref{3a}) can be rewritten
$$
T_{\a_1}(\la \varphi_0\ra,\la \varphi_1\ra,\ldots,\la \varphi_n\ra)
=(\la \varphi_0\ra,\la \varphi_1\ra,\ldots,\la \varphi_n\ra)U'D'L',
$$
where 
$$
U'=J\bar{L}J,\quad D'=J\bar{D}J,\quad L'=J\bar{U}J.
$$
We see that $U'$, $D'$ and $L'$ are upper triangular, diagonal and lower triangular matrices,
respectively with non-zero entries
\begin{eqnarray*}
u'_{ij}&=&(-1)^{j-i}{j\choose i}
\frac{(\a_3+(n-j)\t;\t)_{j-i}}{\big(\a_1+\a_3+2(n-j)\t;\t\big)_{j-i}}
\bigg(\frac{x_3-x_1}{x_2-x_1}\bigg)^{\!\! j-i},
\\[7pt]
d'_{j}&=&\frac{(\a_1;\t)_{n-j}\big(\a_1+\a_3+2(n-j)\t;\t\big)_{j}(x_2-x_1)^{j}(x_3-x_1)^{n-j}}
{(\a_1+\a_3+(n-j-1)\t;\t\big)_{n-j}\big(\a_1+\a_2+\a_3+(2n-j-1)\t;\t\big)_{j}},
\\[7pt]
l'_{ij}&=&
(-1)^{i-j}{n-j\choose n-i}
\frac{
\big(\a_2+j\t;\t\big)_{i-j}}{\big(\a_1+\a_3+2(n-i)\t;\t\big)_{i-j}}.
\end{eqnarray*}

\section{Interpolation polynomials}
In this section we present some lemmas, and collaries of the lemmas, which together imply
Theorem \ref{thm2.2}. We will defer the proof of one of these --- certain key three-term
relations --- until the next section. 
We begin with a lemma which enables the difference
system of Theorem \ref{thm2.2} to be written in a more convenient form.

\begin{lem}\label{lk}
Let $U$ be as in Theorem \ref{thm2.2}. We have that $U^{-1} = (u_{ij}^*)_{0\le i,j\le n}$ is
the upper triangular matrix with non-zero entries 
$$
u_{ij}^*={j\choose i}
\frac{\big(\a_3+(n-j)\t;\t\big)_{j-i}}{\big(\a_1+\a_2+(j+i-1)\t;\t\big)_{j-i}}.
$$
\end{lem}

\noindent
{\bf Proof.} \quad It suffices to check that for $i < j$, $\sum_{k=i}^j u_{ik}
u_{kj}^* = 0$. Now
\begin{eqnarray*}
\lefteqn{\sum_{k=i}^j u_{ik}u_{kj}^*}\\&=&\sum_{k=i}^j
(-1)^{k-i}{k\choose i}
\frac{
\big(\alpha_3+(n-k)\t;\t\big)_{k-i}}{\big(\alpha_1+\alpha_2+2i\t;\t\big)_{k-i}}
{j\choose k}
\frac{\big(\alpha_3+(n-j)\t;\t\big)_{j-k}}{\big(\alpha_1+\alpha_2+(j+k-1)\t;\t\big)_{j-k}}\\
&=&
\frac{\big(\alpha_3+(n-j)\t;\t\big)_{j-i}}{\big(\alpha_1+\alpha_2+2i\t;\t\big)_{2j-2i-1}}
 {j\choose i}\sum_{k=i}^j  (-1)^{k-i}{j-i\choose k-i}\big(\alpha_1+\alpha_2+2i\t+(k-i)\t;\t\big)_{j-i-1}
 \\
&=&
\frac{\big(\alpha_3+(n-j)\t;\t\big)_{j-i}}{\big(\alpha_1+\alpha_2+2i\t;\t\big)_{2j-2i-1}}
 {j\choose i}\sum_{k=0}^{j-i} (-1)^{k} {j-i\choose k}\big(\alpha_1+\alpha_2+2i\t+k\t;\t\big)_{j-i-1},
\end{eqnarray*}
and the last summation vanishes as an example of the summation formula for
${}_2 F_1(a,b;c;1)$. \hfill $\square$

A more convenient form of the difference system can now be established.
\begin{lem}
\label{lem:our result2}
$$
T_{\a_1}(\la \varphi_0\ra,\la \varphi_1\ra,\ldots,\la \varphi_n\ra)
\left(\!\!
\begin{array}{cccc}
\tilde{u}_{00} & \tilde{u}_{01}  & \cdots & \tilde{u}_{0n}\\
  & \tilde{u}_{11}& \cdots & \tilde{u}_{1n}\\
  &   & \cdots & \cdots\\
 & & & \tilde{u}_{nn}
\end{array}
\!\!\right)
=(\la \varphi_0\ra,\la \varphi_1\ra,\ldots,\la \varphi_n\ra)
\left(\!\!
\begin{array}{cccc}
\tilde{l}_{00} &   &  & \\
\tilde{l}_{10} & \tilde{l}_{11} & & \\
\cdots & \cdots & \cdots & \\
\tilde{l}_{n0} & \tilde{l}_{n1} & \cdots & \tilde{l}_{nn}
\end{array}
\!\!\right)
$$
where
\begin{eqnarray*}
\tilde{u}_{ij}&=&{j\choose i}\big(\a_1+\a_2+\a_3+(n+j-1)\t;\t\big)_{n-j}\big(\a_3+(n-j)\t;\t\big)_{j-i}
\big(\a_1+\a_2+(j-1)\t;\t\big)_i,
\\[10pt]
\tilde{l}_{ij}&=&(-1)^{i-j}{n-j\choose n-i}(\a_1;\t)_j(\a_2+j\t;\t)_{i-j}\big(\a_1+\a_2+(i+j)\t;\t\big)_{n-i}(x_3-x_1)^{n-i}(x_2-x_1)^i.
\end{eqnarray*}
\end{lem}

%\noindent
%{\bf Proof.} \quad 
This is obtained by acting on the left of both sides of the difference system of
Theorem \ref{thm2.2} by $U^{-1}$, making use of the explicit form of the latter known from
Lemma \ref{lem:our result2} on the LHS, and clearing denominators. 
%\hfill $\square$

\smallskip
Before considering the proof of this rewrite of the difference system, we make note that in a
special case it implies the recurrence (\ref{Sa}) for the Selberg integral. Thus we note from
(\ref{5}) that with $x_2 = x_3$ we have $\varphi_i(z) = \prod_{i=1}^n (x_2 - z_i)$
independent of $i$. We note too that it follows from the definitions of $\tilde{u}_{ij}$ and
$\tilde{l}_{ij}$ in Lemma \ref{lem:our result2} that
\begin{eqnarray*}
&&\sum_{i=0}^j \tilde{u}_{ij} = \tilde{u}_{00} = (\a_1 + \a_2 + \a_3 + (n-1) \tau;\tau)_n \nonumber \\
&& \sum_{i=j}^n \tilde{l}_{ij} \Big |_{x_2 = x_3} = \tilde{l}_{nn} \Big |_{x_2 = x_3} =
(\a_1;\tau)_n (x_3 - x_1)^n,
\end{eqnarray*}
valid for $j=0,1,\dots,n$, where use has been made of the summation formula for
${}_2 F_1(a,b;c;1)$. It follows that if we set $x_1 = 0$, $x_2 = x_3 = 1$ and replace 
$\alpha_2 + \alpha_3$ with $\alpha_2$ in the difference system of Lemma
\ref{lem:our result2}, then it degenerates to a single equation, which is precisely
(\ref{Sa}).

To derive the difference system of Lemma \ref{lem:our result2}, and thus that of
Theorem \ref{thm2.2}, we introduce the interpolation polynomials $\varphi_{i,j}(z)$ according to
\begin{equation}
\varphi_{i,j}(z):=\underbrace{
(z_1-x_1)(z_2-x_1)\cdots(z_j-x_1)
}_{j}
%\times 
\varphi_i(z)\quad\mbox{for}\quad i,j=0,1,\ldots,n.
\label{eq:eij}
\end{equation}
Note that setting $j=0$ gives $\varphi_{i,0}(x) = \varphi_i(z)$, while setting $j=n$ gives
$$
\varphi_{i,n}(z) = (z_1 - x_1)(z_2 - x_1) \cdots (z_n - x_1) \varphi_i(z), \qquad i=0,1,\dots,n
$$
and so
\begin{equation}\label{7}
T_{\alpha_1} \langle \varphi_i \rangle = \langle \varphi_{i,n} \rangle, \qquad \langle \varphi_i \rangle =
\langle \varphi_{i,0} \rangle.
\end{equation}
Most importantly, the integrals (\ref{4.0}) with $\phi = \varphi_{i+1,j}$, $\varphi_{i,j+1}$,
$\varphi_{i+1,j+1}$, or with $\phi = \varphi_{i,j+1}$, $\varphi_{i,j}$, $\varphi_{i+1,j}$ satisfy certain three-term relations.
%the proof of which will be given in Section 4.

\begin{lem}[Three-term relations]
\label{lem:3term}
For $i,j=0,1,\ldots, n-1$ we have
\begin{eqnarray}
\lefteqn{
\big(\a_1+(n-j-1)\t\big)(x_2-x_1)\la \varphi_{i+1,j}\ra
}\nonumber\\
&=&
\big(\a_3+(n-i-1)\t\big)\la \varphi_{i,j+1}\ra
+\big(\a_1+\a_2+(n+i-j-1)\t\big)\la \varphi_{i+1,j+1}\ra.
\label{eq:up1}
\\
[10pt]
\lefteqn{
\big(\a_1+\a_2+\a_3+(2n-j-2)\t\big)\la \varphi_{i,j+1}\ra
}\nonumber\\
&=&\big(\a_1+\a_2+(n+i-j-1)\t\big)(x_3-x_1)\la \varphi_{i,j}\ra
-(\a_2+ i\t)(x_2-x_1)\la \varphi_{i+1,j}\ra.
\label{eq:down1}
\end{eqnarray}
\end{lem}
\noindent
{\bf Proof.} \quad See Appendix A. \hfill $\Box$
\smallskip

These three-term relations in fact imply the difference system of recurrences in
Lemma \ref{lem:our result2}. To see this we first use an induction on $j$ to deduce from
the three-term relations  
two particular  difference systems.
\begin{cor}
\label{cor:UpDown2}
For $0\le j\le k\le n$ we have
\begin{eqnarray}
\lefteqn{\big(\a_1+(k-j)\t;\t\big)_j(x_2-x_1)^j\la \varphi_{k,n-k}\ra}
\label{eq:up1.5}\\
&=&\sum_{i=0}^j{j\choose i}\big(\a_3+(n-k)\t;\t\big)_{j-i}\big(\a_1+\a_2+(2k-j-1)\t;\t\big)_i\la \varphi_{i+k-j,n-k+j}\ra,
\nonumber\\[10pt]
\lefteqn{\big(\a_1+\a_2+\a_3+(n+j-1)\t;\t\big)_{n-k}\la \varphi_{j,n-j}\ra}
\label{eq:down1.5}\\
&=&\sum_{i=k}^n(-1)^{i-k}{n-k\choose n-i}\big(\a_1+\a_2+(i+2j-k)\t;\t\big)_{n-i}(\a_2+j\t;\t)_{i-k}\nonumber\\
&&\quad\times (x_3-x_1)^{n-i}(x_2-x_1)^{i-k}\la \varphi_{i-k+j,k-j}\ra.
\nonumber
\end{eqnarray}
%\end{cor}
In particular, by setting $k=j$ in the above,
%\begin{cor}
%\label{cor:UpDown2}
for $j=0,1,\ldots, n$ we have
\begin{eqnarray}
\lefteqn{(\a_1;\t)_j(x_2-x_1)^j\la \varphi_{j,n-j}\ra}
\label{eq:up2}\\
&=&\sum_{i=0}^j{j\choose i}\big(\a_3+(n-j)\t;\t\big)_{j-i}\big(\a_1+\a_2+(j-1)\t;\t\big)_i\la \varphi_{i,n}\ra,
\nonumber\\[10pt]
\lefteqn{\big(\a_1+\a_2+\a_3+(n+j-1)\t;\t\big)_{n-j}\la \varphi_{j,n-j}\ra}
\label{eq:down2}\\
&=&\sum_{i=j}^n(-1)^{i-j}{n-j\choose n-i}\big(\a_1+\a_2+(i+j)\t;\t\big)_{n-i}(\a_2+j\t;\t)_{i-j}\nonumber\\
&&\quad\times (x_3-x_1)^{n-i}(x_2-x_1)^{i-j}\la \varphi_{i,0}\ra.
\nonumber
\end{eqnarray}
\end{cor}
\noindent
{\bf Proof.} \quad See Appendix B. \hfill $\Box$

\smallskip
The difference system of Lemma \ref{lem:our result2} can now be derived.

\smallskip

\noindent
{\bf Proof of Lemma \ref{lem:our result2}.} \quad Multiplying (\ref{eq:up2}) and (\ref{eq:down2})
by appropriate factors so as to make their LHS's equal, then equating RHS's gives
$$
\sum_{i=0}^j \tilde{u}_{ij} \langle \varphi_{i,n} \rangle =
\sum_{i=j}^n \tilde{l}_{ij} \langle \varphi_{i,0} \rangle, \qquad j=0,1,\dots,n,
$$
where $\tilde{u}_{ij}$ and $\tilde{l}_{ij}$ are as in Lemma  \ref{lem:our result2}.
Making use of (\ref{7}) this reads
$$
\sum_{i=0}^j \tilde{u}_{ij} T_{\alpha_1} \langle \varphi_i \rangle =
\sum_{i=j}^n \tilde{l}_{ij} \langle \varphi_i \rangle
$$
which is precisely the sought difference system.\hfill $\Box$

\section{Implementing the recurrences}
Consider $\Phi$ specialized to the function $\Phi^*(z)$ of (\ref{5a}) but with $\alpha_1 = \alpha$,
$\alpha_2 = \beta$. Let this be a consequence of setting
\begin{equation}\label{xz0}
\alpha_1 = 1, \quad \alpha_2 = \alpha, \quad \alpha_3 = \beta,
 \quad x_1 = x, \quad x_2 = 0, \quad x_3 =1
\end{equation}
in (\ref{3a}). With this choice of $\Phi^*$ define $\langle \phi \rangle^{\#} = \langle \phi \rangle^*
/ \langle 1 \rangle^*$, where  $\langle \phi \rangle^*$ is specified by (\ref{5b}).
Here our aim is to use the difference system of Theorem \ref{thm2.2} to explictly compute the
polynomial in $x$ specified by
\begin{equation}\label{xz}
\Big \langle \prod_{j=1}^n (x - z_j)^\mu \Big \rangle^{\#}, \qquad \mu \in \mathbb Z_+.
\end{equation}
Let  $(T_{\alpha_1} \langle \varphi_0 \rangle)^{\#}$ refer to
(\ref{T1}) with the substitutions (\ref{xz0}) made afterwards, and normalized
by dividing by $\langle 1 \rangle^*$. Then according to (\ref{T1}), (\ref{5})
\begin{equation}
\label{xzr}
(T_{\alpha_1} \langle \varphi_0 \rangle)^{\#} =   {S_n(\alpha+1,\beta,\tau)
\over S_n(\alpha,\beta,\tau) } \,
\Big ( \Big \langle \prod_{j=1}^n (x - z_j) \Big\rangle^{\#}  \Big |_{\alpha \mapsto
\alpha + 1} \Big ).
\end{equation}
Thus our task is to compute $(T_{\alpha_1}^\mu \langle \varphi_0 \rangle)^{\#}$, as we have
\begin{equation}
\label{xzr1}
(T_{\alpha_1}^\mu \langle \varphi_0 \rangle)^{\#} =  (-1)^{n(\mu - 1)} {S_n(\alpha+1,\beta,\tau)
\over S_n(\alpha,\beta,\tau) } \,
\Big ( \Big \langle \prod_{j=1}^n (x - z_j)^\mu \Big \rangle^{\#} \Big |_{\alpha \mapsto
\alpha + 1} \Big ).
\end{equation}

For $\mu = 1$ the closed form evaluation of (\ref{xz}) is known from the work of 
Aomoto \cite{Ao87} as being
proportional to the Jacobi polynomial $P_N^{(\gamma_1,\gamma_2)}(1 - 2x)$ with
$\gamma_1 = \alpha/\tau - 1$, $\gamma_2 = \beta/\tau - 1$. This in turn can be written in terms of a
Gauss hypergeometric function, giving
\begin{equation}\label{xz1}
\Big \langle \prod_{j=1}^n (x - z_j) \Big \rangle^{\#} = 
\tilde{c} \, {}_2 F_1(-n,(\alpha + \beta)/\tau + n - 1, \alpha/\tau;x),
\end{equation}
where
\begin{equation}
\tilde{c} = {(-1)^n ( \alpha;\tau)_n \over (\alpha + \beta + (n - 1)\tau ;\tau)_n }
\end{equation}
(the factor of $\tilde{c}$ is required to make the coefficient of $x^n$ on the RHS unity). 
We can use knowledge of this to calculate $\langle \varphi_k \rangle^{\#}$ ($k=0,\dots,n$). Once these
have been determined we can use (\ref{9}) with $A =: A_{\alpha_1}$ 
to recursively compute $(T_{\alpha_1}^\mu \langle \varphi_0 \rangle)^{\#}$
according to
\begin{equation}\label{xzp}
\Big (T_{\alpha_1}^\mu (\la \varphi_0\ra,\la \varphi_1\ra,\ldots,\la \varphi_n\ra \Big )^{\#}
=(\la \varphi_0\ra^{\#},\la \varphi_1\ra^{\#},\ldots,\la \varphi_n\ra^{\#})A_1 A_2 \cdots A_\mu
\end{equation}
Upon making use of (\ref{xzr1}) this determines (\ref{xz}). Explicitly, with $(\vec{v})_k$ denoting the
$k$-th component of the row vector $\vec{v}$, we have
\begin{equation}\label{xzp1}
(-1)^{n(\mu-1)} {S_n(\alpha+1,\beta,\tau)
\over S_n(\alpha,\beta,\tau) } \,
\Big ( \Big \langle \prod_{j=1}^n (x - z_j)^\mu \Big \rangle^{\#} \Big ) \Big |_{\alpha \mapsto
\alpha + 1} \Big ) = \Big ( (\la \varphi_0\ra^{\#},\la \varphi_1\ra^{\#},\ldots,\la \varphi_n \ra^{\#})  A_1 A_2 \cdots A_\mu  \Big )_1.
\end{equation}

\begin{lem}\label{LE}
Let $\langle \phi \rangle^{\#}$ be specified as below (\ref{xz0}). We have
\begin{equation}\label{eek}
\langle \varphi_k \rangle^{\#} = (-1)^n {(\alpha;\tau)_n \over (\alpha + \beta + (n-1) \tau;\tau)_n }
{(-\beta - (n-1) \tau ; \tau)_k \over (\alpha;\tau)_k}.
\end{equation}
\end{lem}

\noindent
{\bf Proof.} \quad According to (\ref{T1}), (\ref{5}) and (\ref{xz1})
\begin{equation}\label{xz2}
(T_{\alpha_1} \langle \varphi_0 \rangle)^{\#} = {S_n(\alpha+1,\beta,\tau)
\over S_n(\alpha,\beta,\tau) }
\tilde{c} \, {}_2 F_1\Big (-n,1-\beta/\tau - n, \alpha/\tau;
- {x \over 1 - x} \Big ) \Big |_{\alpha \mapsto \alpha + 1},
\end{equation}
where use has been made of a Kummer relation for ${}_2 F_1$. Note from (\ref{Sa}) that
$$
{S_n(\alpha+1,\beta,\tau)
\over S_n(\alpha,\beta,\tau) } = {(\alpha;\tau)_n \over (\alpha + \beta + (n-1) \tau)_n}.
$$
On the other hand, 
it follows from (\ref{9}) that
\begin{equation}\label{xz3}
(T_{\alpha_1} \langle \varphi_0 \rangle)^{\#} = d_0 \sum_{k=0}^n  \langle \varphi_k \rangle^{\#} l_{k0}
\end{equation}
where, after substituting (\ref{xz0}) in (\ref{9a}),
\begin{equation}
d_0 l_{k0} = {(1 + \alpha; \tau)_n (\alpha; \tau)_k \over (1 + \alpha + \beta + (n-1)\tau; \tau)_n
(1 + \alpha;\tau)_k } \Big ( {n \atop n - k} \Big )
\Big ( {x \over 1 - x} \Big )^k.
\end{equation}
Equating (\ref{xz2}) and (\ref{xz3}) we see that the factor of $(1-x)^n$ cancels, and we can
equate coefficients of $(-x/(1-x))^k$ to deduce (\ref{eek}). \hfill $\square$ 

\medskip
It is also possible to derive (\ref{eek}) independent of knowledge of (\ref{xz1}), using instead
an integration formula in Jack polynomials theory, due to Warnaar \cite{Wa05} (see also \cite{FS09}).
With $\lambda = (\lambda_1,\dots,\lambda_n)$ a partition of non-negative integers, and
$P_\lambda^{(\alpha)}(t) = P_\lambda^{(\alpha)}(t_1,\dots,t_n)$ denoting the symmetric Jack polynomial,
the integration formula reads
\begin{eqnarray}\label{W}
&&\int_{[0,\infty)^n} P_\lambda^{(1/\tau)}(t) \prod_{i=1}^n t_i^{x-1} (1 + t_i)^{-x-y-2(n-1)\tau}
\prod_{1 \le j < k \le n} | t_k - t_j|^{2 \tau} \, dt_1 \cdots dt_n \nonumber \\
&& \qquad  = P_\lambda^{(1/\tau)}(v) \Big |_{v_1 = \cdots = v_n = -1}
{[x + (n-1) \tau]_\lambda^{(1/\tau)} \over [-y+1]_\lambda^{(1/\tau)} }
S_n(x,y,\tau),
\end{eqnarray}
where 
\begin{equation}\label{Wu}
[u]_\kappa^{(\alpha)} = \prod_{j=1}^n {\Gamma(u - (j-1)/\alpha + \kappa_j) \over
\Gamma(u - (j-1)/\alpha) }.
\end{equation}
Our interest is in a transformed version of (\ref{W}).

\begin{cor}
Let $\langle \phi \rangle^{\#}$ be as specified below (\ref{xz0}) and let
$ P_\lambda^{(1/\tau)} ( {1 - z \over z}  ) =  
P_\lambda^{(1/\tau)} ( {1 - z_1 \over z_1},\dots,{1 - z_n \over z_n}  )$. We have
\begin{equation}\label{W1}
\Big \langle P_\lambda^{(1/\tau)}\Big ( {1 - z \over z} \Big ) \Big \rangle^{\#} =
 P_\lambda^{(1/\tau)}(v) \Big |_{v_1 = \cdots = v_n = -1}
{[\beta + (n-1) \tau]_\lambda^{(1/\tau)} \over [-\alpha+1]_\lambda^{(1/\tau)} }.
\end{equation}
\end{cor}

\noindent
{\bf Proof.} \quad This follows by making the change of variables $t_i = (1 - u_i)/u_i$ in 
(\ref{W}), then writing $x=\beta$, $y=\alpha$. \hfill $\square$

\medskip
For $\lambda = 1^k$ (i.e.~1 repeated $k$ times), $k \le n$, we have that
$P_\kappa^{(1/\tau)}(z) = e_k(z)$ where $e_k(z) = e_k(z_1,\dots,z_n)$ denotes
the $k$-th elementary symmetric function. Noting from (\ref{Wu}) that with $\kappa = 1^k$,
$$
[u]_\kappa^{(1/\tau)} = (-1)^k (-u;\tau)_k
$$
we can thus specialize (\ref{W1}) and so reclaim (\ref{eek}).

\begin{cor}
We have
\begin{equation}\label{W2}
\Big \langle e_k \Big ( {1 - z \over z} \Big ) \Big \rangle^{\#} =
\Big ( {n \atop k} \Big ) (-1)^k {(-\beta - (n-1) \tau;\tau)_k \over (\alpha - 1;\tau)_k}.
\end{equation}
This is equivalent to (\ref{eek}).
\end{cor}

\noindent
{\bf Proof.} \quad It remains to explain the final assertion. This is in fact a consequence of the
very definition of $\varphi_k$ as given in (\ref{5}), which gives
\begin{eqnarray}\label{W3}
\langle \varphi_k \rangle^{\#} & = & (-1)^{n-k} \Big \langle \prod_{j=1}^{n-k} z_j \prod_{l=n-k+1}^n (1 - z_l)
\Big \rangle^{\#} \nonumber \\
& = & (-1)^{n-k} \Big ( {n \atop k} \Big )^{-1} {S_{n+1}(\alpha+1,\beta,\tau) \over
S_n(\alpha,\beta,\tau) }
\Big (  \Big \langle e_k \Big ( {1 - z \over z} \Big ) \Big \rangle^{\#} \Big |_{\alpha \mapsto \alpha + 1}
\Big ).
\end{eqnarray}
Substituting (\ref{W2}) in (\ref{W3}) and recalling (\ref{Sa}) reclaims (\ref{eek}).
\hfill $\square$

\medskip
The fact that (\ref{eek}) has been derived independent of (\ref{xz1}) means, from the argument of
the proof of Lemma \ref{LE}, that the difference system (\ref{9}) can be used to prove (\ref{xz1}). 

With (\ref{eek}) substituted in (\ref{xzp1}), and the entries of $A$, (\ref{9a}),
specialized according to (\ref{xz0}), we know all terms in (\ref{xzp1}) except the
average---a polynomial in $x$ of degree $n \mu$---which can therefore by computed by matrix
multiplication. For example, in the case $\alpha = \beta = 2$, $n=5$, $\tau = 5$, this give for 
(\ref{xz})
\begin{eqnarray}\label{ff}
&& {23 \over 5437500} - {23 x \over 65250} + {3197  x^2 \over 261000} - {8993 x^3 \over 56550} + 
  {2117449 x^4 \over 2035800} - {793093 x^5 \over 203580} \nonumber \\
&& \qquad + {601937 x^6 \over 67860} - 
  {4384 x^7 \over 351} + {7457 x^8 \over 702} - 5 x^9 + x^{10}.
\end{eqnarray}
In general (\ref{xz})
in the case $\alpha = \beta$ must be unchanged (up to a possible sign) by the mapping $x = 1-y$.
One can check that (\ref{ff}) has this invariance. 

We have presented the difference system (\ref{5}) both for its theoretical interest, and its
utility in computing the random matrix average (\ref{xz}) in the case $\mu \in \mathbb Z_+$.
Regarding the latter, 
the case $\tau = 1$ of (\ref{xz}), after the
change of variables $z_i = \cos^2 \theta_j/2$, $x=\cos^2 \phi/2$, and for certain values
of $a,b$, corresponds to the $\mu$-th moment of the characteristic polynomial for the classical
groups ${\rm Sp}(2n)$, $O^\pm(2n)$ and $O^\pm(2n+1)$. As such it has appeared in various applied
studies in random matrix theory \cite{FK04,KM04,KO08}.

It was remarked in the Introduction that the differential-recurrence scheme of \cite{Fo93}
can also be used to compute the random matrix average (\ref{xz}) in the case $\mu \in \mathbb Z_+$.
This differential-recurrence is equivalent to
a $(n+1) \times (n+1)$ matrix Fuchsian differential equation \cite{FW07p,Mi07}. The fact that there
is both a differential and difference system for the Selberg correlation integrals is closely related
to there being dynamical difference equations associated with solutions of the
KZ equation \cite{MV02}; indeed as an example of the latter theory the Selberg integral
recurrence (\ref{Sa}) was reclaimed.
In recent years higher order
scalar differential equations have been shown to characterize certain correlation functions
in the two-dimensional Ising model \cite{ZBHM04,ZBHM05,BBGHJMZ09} and certain generating
functions in enumerative combinatorics \cite{GJ06a,JR08}. Finding characterizations of
similar problems, outside the class of averages (\ref{xz}), in terms of higher order
scalar difference equations or matrix difference equations remains an open problem.

Another consequence of our results is in providing a fast method of computation of a certain
class of generalized hypergeometric functions based on Jack polynomials $P_\kappa^{(\alpha)}(z)$.
With $C_\kappa^{(\alpha)}(z) := (\alpha^{|\kappa|} |\kappa|! /d_\kappa') P_\kappa^{(\alpha)}(z)$
denoting a renormalized Jack polynomial (for the definition of the
quantity $d_\kappa'$ see \cite[Eq.~(12.37)]{Fo02}), and $[u]_\kappa^{(\alpha)}$
defined by (\ref{Wu}), the generalized hypergeometric functions of interest are defined by
the infinite series
\begin{equation}\label{2F1}
{}_2 F_1^{(\alpha)}(a_1,a_2;b_1;z) :=
\sum_{\kappa} {1 \over |\kappa|!}
{[a_1]_\kappa^{(\alpha)} [a_2]_\kappa^{(\alpha)} \over [b_1]_\kappa^{(\alpha)}}
C_\kappa^{(\alpha)}(z).
\end{equation}
In the case $n=1$ this reduces to the usual Gauss hypergeometric function. In general, the
computation of this function from the series is an inherently difficult task due to the need
to sum over all partitions $\kappa$ \cite{KE04}. In the special case that $a_1$ is a negative integer
the series terminates and it is equal to a multivariable polynomial. If furthermore 
$z_1=\cdots=z_n = x$ (\ref{2F1}) reduces a polynomial of degree $n |a_1|$ and 
it relates to the average (\ref{xz}) according to \cite[Eq.(13.12)]{Fo02}
\begin{equation}
\Big \langle \prod_{j=1}^n (x - z_j)^\mu \Big \rangle^{\#}
= x^{N\mu} \,
{}_2 F_1^{(\alpha)}(-\mu,\alpha+ (n-1)\tau;\alpha+\beta+2(n-1)\tau;z_1,\dots,z_n)
\Big |_{z_1=\cdots=z_n = 1/x}
\end{equation}
Thus the matrix formula (\ref{xzp1}) can be used to compute this class of ${}_2 F_1^{(\alpha)}$
using O$(n^3)$ operations. In contrast, computation from (\ref{2F1}) requires at least
O($e^{\pi \sqrt{2n/3}}$)  operations, due to the sum over partitions.

\section*{Acknowledgements}
PJF thanks Eric Rains for suggesting this problem. This work was supported by the Australian
Research Council and JSPS Grant-in-Aid for Scientific Research (C) 21540225.

\appendix
\section{Appendix A -- Proof of three-term relations}
Let $\frak{S}_n$ be the symmetric group on $\{1,2,\ldots,n\}$, 
which is generated by the following reflections of the coordinates 
$z=(z_1,z_2,\ldots,z_n)\in {\Bbb C}^n$ :
$$
\sigma_{i,i+1} :(z_1,\ldots,z_i,z_{i+1},\ldots,z_n)\mapsto (z_1,\ldots,z_{i+1},z_i,\ldots,z_n), 
\qquad i=1,2,\ldots,n-1.
$$
For a function $f(z)$ on ${\Bbb C}^n$, 
we define action of the group $\frak{S}_n$ on $f(z)$ by 
$$
(\sigma f)(z):=f(\sigma^{-1}(z))\quad\mbox{for}\quad \sigma\in \frak{S}_n.
$$
We say that a function $f(z)$ on ${\Bbb C}^n$
is {\it symmetric} or {\it skew-symmetric} 
if $\sigma f(z)=f(z)$ or $\sigma f(z)=(\sgn\,\sigma)\,f(z)$ 
for all $\sigma\in \frak{S}_n$, respectively.

In this section, to specify the number $n$ of variables $z_1,z_2,...,z_n$, 
we simply use 
$\varphi_{i,j}^{(n)}(z)$ instead of the polynomials $\varphi_{i,j}(z)$ defined in (\ref{eq:eij}).
The symbol $(n)$ on the right shoulder of $\varphi_{i,j}$ indicates the number of variables of $z=(z_1,z_2,...,z_n)$ for $\varphi_{i,j}(z)$.
\vskip 2mm %

We define the orbit of the polynomial $\varphi_{i,j}^{(n)}(z)$ by 
$$
O_{i,j}^{(n)}:=\{\sigma \varphi_{i,j}^{(n)}(z)\,;\, \sigma\in \frak{S}_n\}, 
$$
and define the sum with respect to the orbit $O_{i,j}^{(n)}$ by 
$$
s_{i,j}^{(n)}(z):=\sum_{\sigma\in \frak{S}_n}\sigma \varphi_{i,j}^{(n)}(z),
$$
which is not always monic.
Since, by definition, we have 
$$
\la \sigma \varphi_{i,j}^{(n)}\ra=\la \varphi_{i,j}^{(n)}\ra \mbox{ for any }\sigma\in \frak{S}_n,
$$
we obtain 
$$
\la s_{i,j}^{(n)}\ra=n!\la  \varphi_{i,j}^{(n)}\ra.
$$
\\

For the point $z=(z_1,z_2,\ldots,z_n)\in {\Bbb C}^{n}$ we set 
$$\widehat{z}_i=(z_1,\ldots,z_{i-1},z_{i+1},\ldots,z_n)\in {\Bbb C}^{n-1} \quad i=1,2,\ldots,n,$$
so that, 
\begin{eqnarray*}
\lefteqn{
\varphi_{i,j}^{(n-1)}(\widehat{z}_n)=
\varphi_{i,j}^{(n-1)}(z_1,z_2,\ldots,z_{n-1})
}\\[5pt]
&=&
\underbrace{(z_1-x_1)\cdots(z_j-x_1)}_j\times\underbrace{(x_2-z_1)\cdots(x_2-z_{n-i-1})}_{n-i-1}
\times\underbrace{ (x_3-z_{n-i})\cdots(x_3-z_{n-1})}_{i}
\end{eqnarray*}
\begin{lem}
\label{lem:nabla4}
Set 
$$
H(z)=H_1(z)+H_2(z)
$$
where
\begin{eqnarray}
&&H_1(z)=
-\sum_{k=1}^n\Big[\a_1(x_2-z_k)(x_3-z_k)+\a_2(x_1-z_k)(x_3-z_k)+\a_3(x_1-z_k)(x_2-z_k)\Big]s_{i,j}^{(n-1)}(\widehat{z}_k),\nonumber\\%[10pt]
\label{eq:H1}\\
&&H_2(z)=\sum_{1\le k<l\le n}\frac{2\t}{z_k-z_l}\Big[(x_1-z_k)(x_2-z_k)(x_3-z_k)s_{i,j}^{(n-1)}(\widehat{z}_k)
\nonumber\\[-4pt]
&&\hskip 8cm 
-(x_1-z_l)(x_2-z_l)(x_3-z_l)s_{i,j}^{(n-1)}(\widehat{z}_l)\Big].
\label{eq:H2}
\end{eqnarray}
Then $\la H\ra=0$.
\end{lem}
{\bf Remark.} Both $H_1(z)$ and $H_2(z)$ are symmetric under the action of $\frak{S}_n$. \\
\par\noindent
{\bf Proof.} Put $\phi(z)=\phi_k(z):=(x_1-z_k)(x_2-z_k)(x_3-z_k)s_{i,j}^{(n-1)}(\widehat{z}_k)$
in (\ref{eq:nabla}) of Lemma \ref{lem:nabla}. 
Since 
$$
\frac{\partial \phi_k}{\partial z_k}=
-\Big[(x_1-z_k)(x_3-z_k)+(x_1-z_k)(x_3-z_k)+(x_1-z_k)(x_2-z_k)\Big]s_{i,j}^{(n-1)}(\widehat{z}_k),
$$
we have $\sum_{k=1}^n \nabla_{\!k}\,\phi_k(z)=H_1(z)+H_2(z)=H(z)$. 
On the other hand, from Lemma \ref{lem:nabla}, we have 
$\la\sum_{k=1}^n \nabla_{\!k}\,\phi_k\ra
=\sum_{k=1}^n \la\nabla_{\!k}\,\phi_k\ra=0$. Therefore $\la H\ra=0$.
\hfill $\Box$
\subsection{Proof of three-term relation (\ref{eq:up1})%Red
}
In this section we will prove (\ref{eq:up1}) in Lemma \ref{lem:3term}. Throughout this section we assume that $n\le i+j$.

First we will show $\la H_1\ra$ is expanded by $\la \varphi_{i+1,j}^{(n)}\ra, \la \varphi_{i,j+1}^{(n)}\ra$ and 
$\la \varphi_{i+1,j+1}^{(n)}\ra$. 
Since 
\begin{eqnarray*}
\lefteqn{
\a_1(x_2-z_k)(x_3-z_k)+\a_2(x_1-z_k)(x_3-z_k)+\a_3(x_1-z_k)(x_2-z_k)}
\\
&=&(\a_1+\a_2)(x_1-z_k)(x_3-z_k)+\a_3(x_1-z_k)(x_2-z_k)+\a_1(x_2-x_1)(x_3-z_k)
\end{eqnarray*}
and $\displaystyle s_{i,j}^{(n-1)}(\widehat{z}_k)=\sum_{\sigma\in \frak{S}_{n-1}}\sigma \varphi_{i,j}^{(n-1)}(\widehat{z}_k)$, 
from (\ref{eq:H1}), $H_1(z)$ is written as
\begin{eqnarray}
H_1(z)&=&
\sum_{k=1}^n\sum_{\sigma\in \frak{S}_{n-1}}
\Big[(\a_1+\a_2)(z_k-x_1)(x_3-z_k)+\a_3(z_k-x_1)(x_2-z_k)\nonumber\\
&&\qquad\qquad\qquad -\a_1(x_2-x_1)(x_3-z_k)\Big]\sigma \varphi_{i,j}^{(n-1)}(\widehat{z}_k).
\label{eq:H1-1}
\end{eqnarray}
Under the assumption $n\le i+j$, we have $n-i-1< j$. Then 
$$
(z_k-x_1)(x_3-z_k)\sigma \varphi_{i,j}^{(n-1)}(\widehat{z}_k)\in O^{(n)}_{i+1,j+1},\quad 
(z_k-x_1)(x_2-z_k)\sigma \varphi_{i,j}^{(n-1)}(\widehat{z}_k)\in O^{(n)}_{i,j+1},
$$
$$
\mbox{ and }
(x_2-x_1)(x_3-z_k)\sigma \varphi_{i,j}^{(n-1)}(\widehat{z}_k)\in O^{(n)}_{i+1,j}.
$$
Since $H_1(z)$ is symmetric, the above implies  
$$
\sum_{k=1}^n\sum_{\sigma\in \frak{S}_{n-1}}(z_k-x_1)(x_3-z_k)\sigma \varphi_{i,j}^{(n-1)}(\widehat{z}_k)
=c\, s_{i+1,j+1}^{(n)}(z),
$$
where $c$ is some integer coefficient. In particular, $c=1$ is confirmed as follows. The LHS is the sum of $n\times (n-1)!$ elements of $O^{(n)}_{i+1,j+1}$, 
while $s_{i+1,j+1}^{(n)}$ in the RHS is the sum of $n!$ elements of $O^{(n)}_{i+1,j+1}$. Thus $n\times (n-1)!=n!\,c$, i.e. $c=1$. In the same way we have 
$$
\sum_{k=1}^n\sum_{\sigma\in \frak{S}_{n-1}}(z_k-x_1)(x_2-z_k)\sigma \varphi_{i,j}^{(n-1)}(\widehat{z}_k)= s_{i,j+1}^{(n)}(z),
$$
$$
\sum_{k=1}^n\sum_{\sigma\in \frak{S}_{n-1}}(x_2-z_k)(x_3-z_k)\sigma \varphi_{i,j}^{(n-1)}(\widehat{z}_k)= s_{i+1,j}^{(n)}(z).
$$
Therefore, from (\ref{eq:H1-1}), we expand $H_1(z)$ as 
\begin{equation}
H_1(z)=(\a_1+\a_2)s_{i+1,j+1}^{(n)}(z)+\a_3 s_{i,j+1}^{(n)}(z)-\a_1(x_2-x_1) s_{i+1,j}^{(n)}(z),
\label{eq:H1-2}
\end{equation}
so that
\begin{equation}
\la H_1\ra
=n!\Big[(\a_1+\a_2)\la \varphi_{i+1,j+1}^{(n)}\ra+\a_3 \la \varphi_{i,j+1}^{(n)}\ra-\a_1(x_2-x_1) \la \varphi_{i+1,j}^{(n)}\ra\Big].
\label{eq:H1red}
\end{equation}
\par
Next we will show $\la H_2\ra$ is expanded by $\la \varphi_{i+1,j}^{(n)}\ra, \la \varphi_{i,j+1}^{(n)}\ra$ and 
$\la \varphi_{i+1,j+1}^{(n)}\ra$. 
We first fix $k,l$ satisfying $1\le k<l\le n$. 
Since, under assumption $n\le i+j$, we have 
\begin{eqnarray*}
\lefteqn{
\varphi_{i,j}^{(n-1)}(z_1,z_2,\ldots,z_{n-1})
}\\[5pt]
&=&\hskip 11.5pt (z_1-x_1)\cdots(z_{n-i-1}-x_1)
\hskip 1pt (z_{n-i}-x_1)\cdots(z_{j}-x_1)\\[-3pt]
&&\times\underbrace{(x_2-z_1)\cdots(x_2-z_{n-i-1})}_{n-i-1}
\underbrace{ (x_3-z_{n-i})\cdots(x_3-z_{j})}_{j-(n-i-1)}
\underbrace{ (x_3-z_{j+1})\cdots(x_3-z_{n-1})}_{n-j-1},
\end{eqnarray*}
if $n\le i+j$, then for each monomial $\sigma \varphi_{i,j}^{(n-1)}(\widehat{z}_k)\in O_{i,j}^{(n-1)}$
one of the following three cases is chosen:  
\begin{itemize}
\item [(1)] $\sigma \varphi_{i,j}^{(n-1)}(\widehat{z}_k)$ has the factor $(z_l-x_1)(x_2-z_l)$,
\item [(2)] $\sigma \varphi_{i,j}^{(n-1)}(\widehat{z}_k)$ has the factor $(z_l-x_1)(x_3-z_l)$,
\item [(3)] $\sigma \varphi_{i,j}^{(n-1)}(\widehat{z}_k)$ has the factor $(x_3-z_l)$, but does not have the factor $(z_l-x_1)$.
\end{itemize}
Here we set $S_i:=\{\sigma\in \frak{S}_{n-1}\,;\, \sigma \mbox{ satisfies the case } (i)\}$ for $i=1,2,3$.    
Since 
the numbers of the factors $(z_l-x_1)(x_2-z_l)$, $(z_l-x_1)(x_3-z_l)$ and $(x_3-z_l)$ 
appearing in $\sigma \varphi_{i,j}^{(n-1)}(\widehat{z}_k)$ are $n-i-1$, $j-(n-i-1)$ and $n-j-1$, respectively, 
we have 
\begin{equation}
\#S_1=(n-i-1)(n-2)!,\ \#S_2=(i+j-n+1)(n-2)!,\ \#S_3=(n-j-1)(n-2)! 
\label{eq:SSS}
\end{equation}
and $\#S_1+\#S_2+\#S_3=\#\frak{S}_{n-1}=(n-1)!$. These numbers are to be used later. 

For each case we will explicitly calculate
$$
\frac{1}{z_k-z_l}\Big[(x_1-z_k)(x_2-z_k)(x_3-z_k)\sigma \varphi_{i,j}^{(n-1)}(\widehat{z}_k)
-(x_1-z_l)(x_2-z_l)(x_3-z_l) \sigma \varphi_{i,j}^{(n-1)}(\widehat{z}_l)\Big].
$$
The point of the calculations below is to use the identities  
$$
\frac{1}{z_k-z_l}\Big[(a-z_k)-(a-z_l)\Big]
=-1
$$
and 
$$
\frac{1}{z_k-z_l}\Big[(a-z_k)(b-z_k)-(a-z_l)(b-z_l)\Big]
=-\Big[(b-a)+(a-z_k)+(a-z_l)\Big].
$$
\underline{Case 1}: We assume that $\sigma \varphi_{i,j}^{(n-1)}(\widehat{z}_k)$ is divisible by $(x_1-z_l)(x_2-z_l)$.  
Then we have 
$$
\frac{\sigma \varphi_{i,j}^{(n-1)}(\widehat{z}_k)}{(x_1-z_l)(x_2-z_l)}
=\frac{\sigma \varphi_{i,j}^{(n-1)}(\widehat{z}_l)}{(x_1-z_k)(x_2-z_k)}.
$$ 
Therefore we obtain 
\begin{eqnarray}
%\lefteqn{
&&\frac{1}{z_k-z_l}\Big[(x_1-z_k)(x_2-z_k)(x_3-z_k)\sigma \varphi_{i,j}^{(n-1)}(\widehat{z}_k)
-(x_1-z_l)(x_2-z_l)(x_3-z_l) \sigma \varphi_{i,j}^{(n-1)}(\widehat{z}_l)\Big]
%}
\nonumber\\
&&=\frac{(x_1-z_k)(x_2-z_k)\sigma \varphi_{i,j}^{(n-1)}(\widehat{z}_k)}{z_k-z_l}\Big[(x_3-z_k)-(x_3-z_l)\Big]
\nonumber\\
&&=(z_k-x_1)(x_2-z_k)\sigma \varphi_{i,j}^{(n-1)}(\widehat{z}_k)\in O_{i,j+1}^{(n)}.
\label{eq:case1}
\end{eqnarray}
\par
\noindent
\underline{Case 2}: We assume that $\sigma \varphi_{i,j}^{(n-1)}(\widehat{z}_k)$ is divisible by $(x_1-z_l)(x_3-z_l)$.  
In the same way as Case 1, we obtain 
\begin{eqnarray*}
&&\frac{1}{z_k-z_l}\Big[(x_1-z_k)(x_2-z_k)(x_3-z_k)\sigma \varphi_{i,j}^{(n-1)}(\widehat{z}_k)
-(x_1-z_l)(x_2-z_l)(x_3-z_l) \sigma \varphi_{i,j}^{(n-1)}(\widehat{z}_l)\Big]
\\
&&=(z_k-x_1)(x_3-z_k)\sigma \varphi_{i,j}^{(n-1)}(\widehat{z}_k)\in O_{i+1,j+1}^{(n)}.
\end{eqnarray*}
\par
\noindent
\underline{Case 3}: We assume that $\sigma \varphi_{i,j}^{(n-1)}(\widehat{z}_k)$ is divisible only by $(x_3-z_l)$.
Since 
$$
\frac{\sigma \varphi_{i,j}^{(n-1)}(\widehat{z}_k)}{(x_3-z_l)}
=\frac{\sigma \varphi_{i,j}^{(n-1)}(\widehat{z}_l)}{(x_3-z_k)},
$$
we obtain 
\begin{eqnarray}
&&\frac{1}{z_k-z_l}\Big[(x_1-z_k)(x_2-z_k)(x_3-z_k)\sigma \varphi_{i,j}^{(n-1)}(\widehat{z}_k)
-(x_1-z_l)(x_2-z_l)(x_3-z_l) \sigma \varphi_{i,j}^{(n-1)}(\widehat{z}_l)\Big]
\nonumber\\
&&=\frac{(x_3-z_k)\sigma \varphi_{i,j}^{(n-1)}(\widehat{z}_k)}{z_k-z_l}\Big[(x_1-z_k)(x_2-z_k)-(x_1-z_l)(x_2-z_l)\Big]\nonumber\\
&&=-(x_3-z_k)\sigma \varphi_{i,j}^{(n-1)}(\widehat{z}_k)\Big[
(x_2-x_1)+(x_1-z_k)+(x_1-z_l)\Big]\nonumber\\
&&=
(z_k-x_1)(x_3-z_k)\sigma \varphi_{i,j}^{(n-1)}(\widehat{z}_k)
+(z_l-x_1)(x_3-z_l)\sigma \varphi_{i,j}^{(n-1)}(\widehat{z}_l)
\nonumber\\
&&\qquad
 -(x_2-x_1)(x_3-z_k)\sigma \varphi_{i,j}^{(n-1)}(\widehat{z}_k).
\label{eq:case3}
\end{eqnarray}
Note that    
$(x_3-z_k)\sigma \varphi_{i,j}^{(n-1)}(\widehat{z}_k)\in O_{i+1,j}^{(n)}$
and
$$(x_1-z_k)(x_3-z_k)\sigma \varphi_{i,j}^{(n-1)}(\widehat{z}_k),\ 
(x_1-z_l)(x_3-z_l)\sigma \varphi_{i,j}^{(n-1)}(\widehat{z}_l)\in O_{i+1,j+1}^{(n)}.$$

Considering the above cases, since 
$H_2(z)$ is symmetric, $H_2(z)/\tau$ is expanded as 
\begin{equation}
\frac{H_2(z)}{\tau}=c_1 s_{i+1,j+1}^{(n)}(z)+
c_2 s_{i,j+1}^{(n)}(z)-(x_2-x_1)c_3 s_{i+1,j}^{(n)}(z)
\label{eq:H2/t-1}
\end{equation}
where $c_1,c_2$ and $c_3$ are some integer coefficients. On the other hand 
$H_2(z)/\tau$ is also expanded as
\begin{eqnarray}
\frac{H_2(z)}{\tau}&=&2\sum_{1\le k<l\le n}
\bigg[ 
\sum_{\sigma\in S_1}(z_k-x_1)(x_2-z_k)\sigma \varphi_{i,j}^{(n-1)}(\widehat{z}_k)
+\sum_{\sigma\in S_2}(z_k-x_1)(x_3-z_k)\sigma \varphi_{i,j}^{(n-1)}(\widehat{z}_k)\nonumber\\
&&
\quad\qquad\qquad
+\sum_{\sigma\in S_3}\Big((x_1-z_k)(x_3-z_k)\sigma \varphi_{i,j}^{(n-1)}(\widehat{z}_k)+
(x_1-z_l)(x_3-z_l)\sigma \varphi_{i,j}^{(n-1)}(\widehat{z}_l)\nonumber\\
&&\qquad\qquad\qquad\qquad
-(x_2-x_1)(x_3-z_k)\sigma \varphi_{i,j}^{(n-1)}(\widehat{z}_k)\Big)
\bigg].
\label{eq:H2/t-2}
\end{eqnarray}
The RHS of (\ref{eq:H2/t-2}) includes the sum of $2{n \choose 2}(\#S_2+2\#S_3)$ elements of $O_{i+1,j+1}^{(n)}$,
while $s_{i+1,j+1}^{(n)}(z)$ in the RHS of (\ref{eq:H2/t-1}) is the sum of $n!$ elements of $O_{i+1,j+1}^{(n)}$. 
Thus we have $n!\,c_1=2{n \choose 2}(\#S_2+2\#S_3)$. In the same way we have  
$n!\, c_2=2{n \choose 2}\#S_1$ and $n!\, c_3=2{n \choose 2}\#S_3$. From (\ref{eq:SSS}), we obtain 
$$
c_1=n-j+i-1, \quad c_2=n-i-1 \quad\mbox{ and }\quad c_3=n-j-1.
$$
Then (\ref{eq:H2/t-1}) implies 
\begin{eqnarray}
\la H_2\ra
=n!\t
\Big[(n-j+i-1)\la \varphi_{i+1,j+1}^{(n)}\ra
%\nonumber\\
%&&\hskip 2cm 
+(n-i-1)\la \varphi_{i,j+1}^{(n)}\ra-(n-j-1)(x_2-x_1)\la \varphi_{i+1,j}^{(n)}\ra\Big ].
%\nonumber\\
\label{eq:H2red}
\end{eqnarray}
From (\ref{eq:H1red}), (\ref{eq:H2red}) and Lemma \ref{lem:nabla4}, i.e., $\la H_2\ra+\la H_1\ra=\la H\ra=0$, we therefore obtain (\ref{eq:up1}). 
\hfill $\Box$

\subsection{Proof of three-term relation (\ref{eq:down1})%Black
}
In this section we will prove (\ref{eq:down1}) in Lemma \ref{lem:3term}. Throughout this section we assume that $i+j\le n$.

First we will show $\la H_1\ra$ is expanded by $\la \varphi_{i,j}^{(n)}\ra, \la \varphi_{i+1,j}^{(n)}\ra$ and 
$\la \varphi_{i,j+1}^{(n)}\ra$. 
Since 
\begin{eqnarray*}
\lefteqn{
\a_1(x_2-z_k)(x_3-z_k)+\a_2(x_1-z_k)(x_3-z_k)+\a_3(x_1-z_k)(x_2-z_k)}
\\
&=&-(\a_1+\a_2+\a_3)(z_k-x_1)(x_2-z_k)+(\a_1+\a_2)(x_3-x_1)(x_2-z_k)-\a_2(x_2-x_1)(x_3-z_k)
\end{eqnarray*}
and (\ref{eq:H1}), 
as we gave the expression (\ref{eq:H1-2}) of $H_1(z)$ in the previous section, 
we can expand $H_1(z)$ as 
$$
H_1(z)=(\a_1+\a_2+\a_3) s_{i,j+1}^{(n)}(z)-(\a_1+\a_2)(x_3-x_1)s_{i,j}^{(n)}(z)+\a_2(x_2-x_1)s_{i+1,j}^{(n)}(z),
$$
so that
\begin{equation}
\la H_1\ra=n!\Big[(\a_1+\a_2+\a_3) \la \varphi_{i,j+1}^{(n)}\ra-(\a_1+\a_2)(x_3-x_1)\la \varphi_{i,j}^{(n)}\ra
+\a_2(x_2-x_1) \la \varphi_{i+1,j}^{(n)}\ra\Big].
\label{eq:H1black}
\end{equation}
\par
Next we will show $\la H_2\ra$ is expanded by $\la \varphi_{i,j}^{(n)}\ra, \la \varphi_{i+1,j}^{(n)}\ra$ and 
$\la \varphi_{i,j+1}^{(n)}\ra$. 
We first fix $k,l$ satisfying $1\le k<l\le n$. 
Since, under assumption $i+j\le n$, we have 
\begin{eqnarray*}
\lefteqn{
\varphi_{i,j}^{(n-1)}(z_1,z_2,\ldots,z_{n-1})
}\\[5pt]
&=&\hskip 11.5pt (z_1-x_1)\cdots(z_{j}-x_1)\\[-3pt]
&&\times\underbrace{(x_2-z_1)\cdots(x_2-z_{j})}_{j}
\underbrace{ (x_2-z_{j+1})\cdots(x_2-z_{n-i})}_{(n-j-1)-i}
\underbrace{ (x_3-z_{n-i})\cdots(x_3-z_{n-1})}_{i},
\end{eqnarray*}
if $i+j\le n$, then for each monomial $\sigma \varphi_{i,j}^{(n-1)}(\widehat{z}_k)\in O_{i,j}^{(n-1)}$
one of the following three cases is chosen:  
\begin{itemize}
\item [(1)] $\sigma \varphi_{i,j}^{(n-1)}(\widehat{z}_k)$ has the factor $(z_l-x_1)(x_2-z_l)$,
\item [(2)] $\sigma \varphi_{i,j}^{(n-1)}(\widehat{z}_k)$ has the factor $(x_2-z_l)$, but does not have the factor $(z_l-x_1)$,
\item [(3)] $\sigma \varphi_{i,j}^{(n-1)}(\widehat{z}_k)$ has the factor $(x_3-z_l)$.
\end{itemize}
%
%\par
%\noindent
\underline{Case 1}: We assume that $\sigma \varphi_{i,j}^{(n-1)}(\widehat{z}_k)$ is divisible by $(x_1-z_l)(x_2-z_l)$. 
In the same way as the previous section, from (\ref{eq:case1}), we have 
\begin{eqnarray*}
&&\frac{1}{z_k-z_l}\Big[(x_1-z_k)(x_2-z_k)(x_3-z_k)\sigma \varphi_{i,j}^{(n-1)}(\widehat{z}_k)
-(x_1-z_l)(x_2-z_l)(x_3-z_l) \sigma \varphi_{i,j}^{(n-1)}(\widehat{z}_l)\Big]
\nonumber\\
&&=(z_k-x_1)(x_2-z_k)\sigma \varphi_{i,j}^{(n-1)}(\widehat{z}_k)\in O_{i,j+1}^{(n)}.
\end{eqnarray*}
\par
\noindent
\underline{Case 2}: We assume that $\sigma \varphi_{i,j}^{(n-1)}(\widehat{z}_k)$ is divisible only by $(x_2-z_l)$.
In the same way as the previous section, from (\ref{eq:case3}), we have
\begin{eqnarray*}
&&\frac{1}{z_k-z_l}\Big[(x_1-z_k)(x_2-z_k)(x_3-z_k)\sigma \varphi_{i,j}^{(n-1)}(\widehat{z}_k)
-(x_1-z_l)(x_2-z_l)(x_3-z_l) \sigma \varphi_{i,j}^{(n-1)}(\widehat{z}_l)\Big]
\\
&&=
(z_k-x_1)(x_2-z_k)\sigma \varphi_{i,j}^{(n-1)}(\widehat{z}_k)
+(z_l-x_1)(x_2-z_l)\sigma \varphi_{i,j}^{(n-1)}(\widehat{z}_l)
%\\
%&&\qquad
 -(x_3-x_1)(x_2-z_k)\sigma \varphi_{i,j}^{(n-1)}(\widehat{z}_k).
\end{eqnarray*}
Note that    
$(x_2-z_k)\sigma \varphi_{i,j}^{(n-1)}(\widehat{z}_k)\in O_{i,j}^{(n)}$
and
$$(x_1-z_k)(x_2-z_k)\sigma \varphi_{i,j}^{(n-1)}(\widehat{z}_k),\ 
(x_1-z_l)(x_2-z_l)\sigma \varphi_{i,j}^{(n-1)}(\widehat{z}_l)\in O_{i,j+1}^{(n)}.$$
\par
\noindent
\underline{Case 3}: We assume that $\sigma \varphi_{i,j}^{(n-1)}(\widehat{z}_k)$ is divisible by $(x_3-z_l)$.
Since 
$$
(x_1-z_k)(x_3-z_k)=(x_1-z_k)(x_2-z_k)-(x_2-x_1)(x_3-z_k)+(x_3-x_1)(x_2-z_k),
$$
we have 
\begin{eqnarray}
\lefteqn{(z_k-x_1)(x_3-z_k)\sigma \varphi_{i,j}^{(n-1)}(\widehat{z}_k)}
\nonumber\\
&=&(z_k-x_1)(x_2-z_k)\sigma \varphi_{i,j}^{(n-1)}(\widehat{z}_k)
+(x_2-x_1)(x_3-z_k)\sigma \varphi_{i,j}^{(n-1)}(\widehat{z}_k)
\nonumber\\&&
-(x_3-x_1)(x_2-z_k)\sigma \varphi_{i,j}^{(n-1)}(\widehat{z}_k).
\label{eq:riemannlike}
\end{eqnarray}
Combining (\ref{eq:case3}) and (\ref{eq:riemannlike}), we obtain 
\begin{eqnarray*}
&&\frac{1}{z_k-z_l}\Big[(x_1-z_k)(x_2-z_k)(x_3-z_k)\sigma \varphi_{i,j}^{(n-1)}(\widehat{z}_k)
-(x_1-z_l)(x_2-z_l)(x_3-z_l) \sigma \varphi_{i,j}^{(n-1)}(\widehat{z}_l)\Big]
\\
&&=
(z_k-x_1)(x_3-z_k)\sigma \varphi_{i,j}^{(n-1)}(\widehat{z}_k)
+(z_l-x_1)(x_3-z_l)\sigma \varphi_{i,j}^{(n-1)}(\widehat{z}_l)
\\
&&\quad
 -(x_2-x_1)(x_3-z_k)\sigma \varphi_{i,j}^{(n-1)}(\widehat{z}_k)\qquad \mbox{(from Eq.(\ref{eq:case3}))}\\
&&=
(z_k-x_1)(x_2-z_k)\sigma \varphi_{i,j}^{(n-1)}(\widehat{z}_k)
-(x_3-x_1)(x_2-z_k)\sigma \varphi_{i,j}^{(n-1)}(\widehat{z}_k)\\
&&\qquad+(z_l-x_1)(x_2-z_l)\sigma \varphi_{i,j}^{(n-1)}(\widehat{z}_l)
+(x_2-x_1)(x_3-z_l)\sigma \varphi_{i,j}^{(n-1)}(\widehat{z}_l)\\
&&\qquad-(x_3-x_1)(x_2-z_l)\sigma \varphi_{i,j}^{(n-1)}(\widehat{z}_l)
\qquad \mbox{(from Eq.(\ref{eq:riemannlike}))}.
\end{eqnarray*}
Here 
$$
(z_k-x_1)(x_2-z_k)\sigma \varphi_{i,j}^{(n-1)}(\widehat{z}_k),\ 
(z_k-x_1)(x_2-z_l)\sigma \varphi_{i,j}^{(n-1)}(\widehat{z}_l)\in O_{i,j+1}^{(n)},
$$
$$
(x_2-z_k)\sigma \varphi_{i,j}^{(n-1)}(\widehat{z}_k),\  (x_2-z_l)\sigma \varphi_{i,j}^{(n-1)}(\widehat{z}_l)\in O_{i,j}^{(n)},
$$
$$
(x_3-z_l)\sigma \varphi_{i,j}^{(n-1)}(\widehat{z}_l)\in O_{i+1,j}^{(n)}.
$$

Considering the above cases, since 
the numbers of the factors $(z_l-x_1)(x_2-z_l)$, $(x_2-z_l)$ and $(x_3-z_l)$ 
appearing in $\sigma \varphi_{i,j}^{(n-1)}(\widehat{z}_k)$ are $j$, $(n-j-1)-i$ and $i$, respectively, 
as we saw in the previous section,  
the function $H_2(z)$ is expanded in total as 
\begin{eqnarray*}
H_2(z)&=&
\frac{2\t{n\choose 2}(n-2)!}{n!}
\Big[js_{i,j+1}^{(n)}(z)+(n-i-j-1)\Big(2s_{i,j+1}^{(n)}(z)-(x_3-x_1)s_{i,j}^{(n)}(z)\Big)
\nonumber\\
&&\hskip 3.0cm +i\Big(2s_{i,j+1}^{(n)}(z)-2(x_3-x_1)s_{i,j}^{(n)}(z)
 +(x_2-x_1)s_{i+1,j}^{(n)}(z)\Big)\Big ]
\nonumber\\
&=&\t
\Big[(2n-j-2)s_{i,j+1}^{(n)}(z)
%\nonumber\\
%&&\hskip 2cm 
-(n+i-j-1)(x_3-x_1)s_{i,j}^{(n)}(z)+i(x_2-x_1)s_{i+1,j}^{(n)}(z)\Big],
\end{eqnarray*}
so that 
\begin{eqnarray}
\la H_2\ra
=n!\t
\Big[(2n-j-2)\la \varphi_{i,j+1}^{(n)}\ra
%\nonumber\\
%&&\hskip 2cm 
-(n+i-j-1)(x_3-x_1)\la \varphi_{i,j}^{(n)}\ra+i(x_2-x_1)\la \varphi_{i+1,j}^{(n)}\ra\Big ].
%\nonumber\\
\label{eq:H2black}
\end{eqnarray}
From (\ref{eq:H1black}), (\ref{eq:H2black}) and Lemma \ref{lem:nabla4}, i.e., $\la H_2\ra+\la H_1\ra=\la H\ra=0$, we therefore obtain (\ref{eq:down1}). \hfill $\Box$
\section{Appendix B -- Proof of Corollary \ref{cor:UpDown2}}
We detail the case of Eq.\,(\ref{eq:up1.5}) only; the proof of
(\ref{eq:down1.5}) proceeds similarly.\\
\underline{Induction on $j$}:  
If $j=1$, then Eq.\,(\ref{eq:up1.5}) reads 
$$
\big(\a_1+(k-1)\t\big)(x_2-x_1)\la \varphi_{k,n-k}\ra
=\big(\a_3+(n-k)\t\big)\la \varphi_{k-1,n-k+1}\ra+\big(\a_1+\a_2+(2k-2)\t\big)\la \varphi_{k,n-k+1}\ra,
$$
which is confirmed from (\ref{eq:up1}) by setting $i=k-1$ and $j=n-k$.
Next we assume the following as an inductive hypothesis
\begin{eqnarray}
\lefteqn{\big(\a_1+(k-j+1)\t;\t\big)_{j-1}(x_2-x_1)^{j-1}\la \varphi_{k,n-k}\ra}\label{eq:up3}\\
&=&\sum_{i=0}^{j-1}{j-1\choose i}\big(\a_3+(n-k)\t;\t\big)_{j-i-1}\big(\a_1+\a_2+(2k-j)\t;\t\big)_i
\la \varphi_{i+k-j+1,n-k+j-1}\ra.
\nonumber
\end{eqnarray}
On the other hand, from (\ref{eq:up1}) by setting $i\to i+k-j$ and $j\to n-k+j-1$ we have
\begin{eqnarray}
\lefteqn{\big(\a_1+(k-j)\t\big)(x_2-x_1)\la \varphi_{i+k-j+1,n-k+j-1}\ra}\label{eq:up4}\\
&=&\big(\a_3+(n-i-k+j-1)\t\big)\la \varphi_{i+k-j,n-k+j}\ra+\big(\a_1+\a_2+(i+2k-2j)\t\big)\la \varphi_{i+k-j+1,n-k+j}\ra.
\nonumber
\end{eqnarray}
Then the LHS of (\ref{eq:up1.5}) is written as 
\begin{eqnarray}
\lefteqn{\big(\a_1+(k-j)\t;\t\big)_j(x_2-x_1)^j\la \varphi_{k,n-k}\ra}\nonumber\\
&=&\big(\a_1+(k-j)\t\big)(x_2-x_1)
\nonumber\\
&&\times
\sum_{i=0}^{j-1}{j-1\choose i}\big(\a_3+(n-k)\t;\t\big)_{j-i-1}\big(\a_1+\a_2+(2k-j)\t;\t\big)_i
\la \varphi_{i+k-j+1,n-k+j-1}\ra
\nonumber\\
&&\hskip 106mm\mbox{(from (\ref{eq:up3}))}
\nonumber\\
&=&\sum_{i=0}^{j-1}{j-1\choose i}\big(\a_3+(n-k)\t;\t\big)_{j-i-1}\big(\a_1+\a_2+(2k-j)\t;\t\big)_i
\nonumber\\
&&\quad\times \Big[\big(\a_3+(n-i-k+j-1)\t\big)\la \varphi_{i+k-j,n-k+j}\ra
\nonumber\\
&&\qquad\qquad
+\big(\a_1+\a_2+(i+2k-2j)\t\big)\la \varphi_{i+k-j+1,n-k+j}\ra\Big]
\qquad\mbox{(from (\ref{eq:up4}))}
\nonumber
\\
%
%\end{eqnarray}
%\begin{eqnarray}
%
%
&=&\sum_{i=0}^{j-1}{j-1\choose i}\big(\a_3+(n-k)\t;\t\big)_{j-i-1}\big(\a_1+\a_2+(2k-j)\t;\t\big)_i\nonumber\\
&&\qquad\times
\big(\a_3+(n-i-k+j-1)\t\big)\la \varphi_{i+k-j,n-k+j}\ra
\nonumber\\
&&+\sum_{i=1}^{j}{j-1\choose i-1}\big(\a_3+(n-k)\t;\t\big)_{j-i}\big(\a_1+\a_2+(2k-j)\t;\t\big)_{i-1}\nonumber\\
&&\qquad\times
\big(\a_1+\a_2+(i+2k-2j-1)\t\big)\la \varphi_{i+k-j,n-k+j}\ra
\nonumber\\
&=&\sum_{i=0}^{j}
\big(\a_3+(n-k)\t;\t\big)_{j-i}\big(\a_1+\a_2+(2k-j)\t;\t\big)_{i-1}
\nonumber\\
&&\times\bigg[{j-1\choose i}\big(\a_1+\a_2+(2k-j+i-1)\t\big)
+{j-1\choose i-1}\big(\a_1+\a_2+(i+2k-2j-1)\t\big)\bigg]
\nonumber\\
&&\times
\la \varphi_{i+k-j,n-k+j}\ra.
\label{eq:up5}
\end{eqnarray}
Here
\begin{eqnarray}
\lefteqn{{j-1\choose i}\big(\a_1+\a_2+(2k-j+i-1)\t\big)
+{j-1\choose i-1}\big(\a_1+\a_2+(i+2k-2j-1)\t\big)}
\nonumber\\
&=&\bigg[{j-1\choose i}
+{j-1\choose i-1}\bigg]\big(\a_1+\a_2+(2k-j+i-1)\t\big)
-j{j-1\choose i-1}\t
\nonumber\\
&=&{j\choose i}\big(\a_1+\a_2+(2k-j+i-1)\t\big)
-i{j\choose i}\t
={j\choose i}\big(\a_1+\a_2+(2k-j-1)\t\big).
\label{eq:up6}
\end{eqnarray}
From (\ref{eq:up5}) and (\ref{eq:up6}) we obtain Eq.\,(\ref{eq:up1.5}).

%\bibliographystyle{amsplain}
%\bibliography{book1}
\providecommand{\bysame}{\leavevmode\hbox to3em{\hrulefill}\thinspace}
\providecommand{\MR}{\relax\ifhmode\unskip\space\fi MR }
% \MRhref is called by the amsart/book/proc definition of \MR.
\providecommand{\MRhref}[2]{%
  \href{http://www.ams.org/mathscinet-getitem?mr=#1}{#2}
}
\providecommand{\href}[2]{#2}

\end{document}